\documentclass[twocolumn,twocolappendix]{aastex631}

\newcommand{\sref}{\S\ref} 

\definecolor{notes}{HTML}{C70039}
\definecolor{softgreen}{HTML}{468465}

\newcommand{\tay}[1]{{#1}}

\newcommand{\hb}{\hbox{\sc H$\beta$}}       
\newcommand{\oiii}{\hbox{\sc [O\,iii]}}     

\newcommand{\jwst}{{\it JWST}}

\newcommand{\jdrp}{\texttt{jwst}}

\newcommand{\lam}{$\lambda$}
\newcommand{\unit}[1]{\ensuremath{\mathrm{\,#1}}\xspace}

\newcommand{\e}{\unit{e^{-}}}

\mathchardef\mhyphen="2D

\newlength{\dhatheight}

\graphicspath{{./}{figures/}}

\begin{document}

\title{TEMPLATES: A Robust Outlier Rejection Method for \jwst/NIRSpec Integral Field Spectroscopy}

\author[0000-0001-6251-4988]{Taylor A. Hutchison}
\email{taylor.hutchison@nasa.gov}
\altaffiliation{NASA Postdoctoral Fellow}
\affiliation{Astrophysics Science Division, Code 660, NASA Goddard Space Flight Center, 8800 Greenbelt Rd, Greenbelt, MD 20771, USA}

\author[0000-0003-1815-0114]{Brian D. Welch}
\affiliation{Astrophysics Science Division, Code 660, NASA Goddard Space Flight Center, 8800 Greenbelt Rd, Greenbelt, MD 20771, USA}
\affiliation{Department of Astronomy, University of Maryland, College Park, MD 20742, USA}
\affiliation{Center for Research and Exploration in Space Science and Technology, NASA/GSFC, Greenbelt, MD 20771 USA}

\author[0000-0002-7627-6551]{Jane R. Rigby}
\affiliation{Astrophysics Science Division, Code 660, NASA Goddard Space Flight Center, 8800 Greenbelt Rd, Greenbelt, MD 20771, USA}

\author[0000-0002-4606-4240]{Grace M. Olivier}
\affiliation{Department of Physics and Astronomy and George P. and Cynthia Woods Mitchell Institute for Fundamental Physics and Astronomy, Texas A\&M University, 4242 TAMU, College Station, TX 77843-4242, US}

\author[0000-0002-3272-7568]{Jack E. Birkin}
\affiliation{Department of Physics and Astronomy and George P. and Cynthia Woods Mitchell Institute for Fundamental Physics and Astronomy, Texas A\&M University, 4242 TAMU, College Station, TX 77843-4242, US}

\author[0000-0001-7946-557X]{Kedar A. Phadke}
\affiliation{Department of Astronomy, University of Illinois, 1002 West Green St., Urbana, IL 61801, USA}
\affiliation{Center for AstroPhysical Surveys, National Center for Supercomputing Applications, 1205 West Clark Street, Urbana, IL 61801, USA}

\author[0000-0002-3475-7648]{Gourav Khullar}
\affiliation{Department of Physics and Astronomy and PITT PACC, University of Pittsburgh, Pittsburgh, PA 15260, USA}

\author[0000-0003-2662-6821]{Bernard J. Rauscher}
\affiliation{Astrophysics Science Division, NASA Goddard Space Flight Center, 8800 Greenbelt Rd, Greenbelt, MD 20771, USA}

\author[0000-0002-7559-0864]{Keren Sharon}
\affiliation{Department of Astronomy, University of Michigan, 1085 S. University Ave, Ann Arbor, MI 48109, USA}

\author[0000-0002-6290-3198]{Manuel Aravena}
\affiliation{N\'{u}cleo de Astronom\'{i}a, Facultad de Ingenier\'{i}a y Ciencias, Universidad Diego Portales, Av. Ej\'{e}rcito Libertador 441, Santiago, Chile}

\author[0000-0003-1074-4807]{Matthew B. Bayliss}
\affiliation{Department of Physics, University of Cincinnati, Cincinnati, OH 45221, USA}

\author[0009-0007-6157-7398]{Lauren A. Elicker}
\affiliation{Department of Physics, University of Cincinnati, Cincinnati, OH 45221, USA}

\author[0000-0002-6787-3020]{Seonwoo Kim}
\affiliation{Department of Astronomy, University of Illinois, 1002 West Green St., Urbana, IL 61801, USA}

\author[0000-0001-6629-0379]{Manuel Solimano}
\affiliation{N\'{u}cleo de Astronom\'{i}a, Facultad de Ingenier\'{i}a y Ciencias, Universidad Diego Portales, Av. Ej\'{e}rcito Libertador 441, Santiago, Chile}

\author[0000-0001-7192-3871]{Joaquin D. Vieira}
\affiliation{Department of Astronomy, University of Illinois, 1002 West Green St., Urbana, IL 61801, USA}
\affiliation{Center for AstroPhysical Surveys, National Center for Supercomputing Applications, 1205 West Clark Street, Urbana, IL 61801, USA}

\author[0000-0002-0786-7307]{David Vizgan}
\affiliation{Department of Astronomy, University of Illinois, 1002 West Green St., Urbana, IL 61801, USA}


\collaboration{50}{on behalf of the \jwst\ TEMPLATES Early Release Science Team}

\begin{abstract}

We describe a custom outlier rejection algorithm for \jwst/NIRSpec integral field spectroscopy.  This method uses a layered sigma clipping approach that adapts clipping thresholds based upon 
the spatial profile of the science target.  We find that this algorithm produces a robust outlier rejection while simultaneously preserving the signal of the science target. Originally developed as a response to unsatisfactory initial performance of the \jdrp\ pipeline outlier detection step, this method works either as a standalone solution, or as a supplement to the current pipeline software.
Comparing leftover (i.e., not flagged) artifacts with the current pipeline's outlier detection step, we find that our method results in one fifth as many residual artifacts as the \jdrp\ pipeline. However, we find a combination of both methods removes nearly all artifacts --- an approach that takes advantage of both our algorithm's robust outlier rejection and the pipeline's use of individual dithers. This combined approach is what the TEMPLATES Early Release Science team has converged upon for our NIRSpec observations.
Finally, we publicly release the code and Jupyter notebooks for the custom outlier rejection algorithm.

\end{abstract}


\section{Introduction} \label{sec:intro}

One of the transformative new capabilities of \jwst\ \citep{Gardner.2023,Rigby.2023} is integral field spectroscopy (IFS) that captures spatially-resolved spectra of distant galaxies with the NIRSpec \citep{Boker.2023} and MIRI \citep{Wright.2023} science instruments.  In the first 1.5 years of \jwst\ science, the user community has had to learn best practices for calibrating and removing instrumental signatures from these data, including expected issues like cosmic rays and surprising issues like residual pattern noise \citep{Rauscher.2023}.
Indeed, the Early Release Science (ERS) teams were charged with testing existing data processing software (such as the STScI \jdrp\ pipeline; \citealt{Bushouse.2023})\footnote{https://jwst-pipeline.readthedocs.io/} and developing resources to work around problems in the existing data reduction pipeline.

One such problem arose in a critical step in the final stage of the \jdrp\ calibration pipeline: the outlier rejection step.  In this step, the pipeline checks for and removes artifacts (outliers, primarily from cosmic rays) present in the individual dithers such that the final three-dimensional (3D) cube is clean of these features.  However, during the first year of observations, this step did not work --- aggressively removing outliers to the extent that the step also removed real flux from the science targets \citep[\jdrp\ calibration pipeline versions 1.10.2 and earlier; e.g.,][]{Perna.2023,Veilleux.2023,Marshall.2023,Vayner.2023}.  \tay{In response, early teams developed alternative processing for this critical step \citep[e.g., running {\sc lacosmic}, \citealt{vanDokkum.2001}, on individual dithers and post-processing the final cube using uniform sigma clipping; see][among others]{Perna.2023}.}    

Part of the challenge of working with early science data from \jwst\ has been adapting to rapid changes in the \jdrp\ pipeline algorithms and calibration files.  One year into the \jwst\ science mission, the pipeline and calibration files were updated such that the outlier rejection step no longer removes flux from the science targets (\jdrp\ calibration pipeline versions 1.11.3 and later).  
However, \tay{crucially,} the pipeline algorithm is still not catching some outliers.  

In this paper, we present a custom outlier rejection algorithm designed to robustly clip the numerous outliers present in the NIRSpec integral field unit (IFU) data, which are largely produced by cosmic rays.  This custom algorithm employs a layered sigma clipping treatment of the 3D data cube, using the signal-to-noise (S/N) spatial profile of the science target \tay{to clip outliers within S/N layers at each wavelength slice}.  This approach removes outliers while preserving the signal of the science targets, a method that is effective even in the presence of very bright emission lines.

We test the effectiveness of a) our algorithm alone, b) the current 
\jdrp\ pipeline outlier detection step, and c) a hybrid approach where we use the \jdrp\ pipeline \textit{and} our custom outlier rejection algorithm to produce results that are superior to the pipeline's alone while taking advantage of the pipeline's use of dithers. 

This paper is organized as follows.
In \S\ref{sec:algorithm} we describe the process of generating a pixel mask to identify the spaxels associated with the science target, defining mask layers for use in the custom outlier rejection algorithm, and the three-part outlier rejection algorithm itself.  In \S\ref{sec:analysis}, we discuss the results of this algorithm on \jwst\ data from the Early Release Science (ERS) program TEMPLATES (PID: 1355; PI Rigby, Co-PI Vieira) and analyze comparisons between this algorithm and the updated pipeline outlier rejection step.  We summarize this work in \S\ref{sec:conclude}, and highlight alternative potential uses and approaches for this algorithm.

\section{The Algorithm} \label{sec:algorithm}

The algorithm described in this work is a custom outlier rejection method designed to robustly flag and remove the numerous outliers present in \jwst/NIRSpec IFU data.  
\tay{We note that the algorithm does require some manual customization, but it produces one of the lowest rates of leftover outliers in the resulting data cubes. Our method works as follows:}
\tay{\begin{itemize}
    \item Sorts spatial pixels as science target pixels or not, based upon the S/N of some spectral feature in the science target.
    \item Separates science target into layers (bins) of S/N.
    \item Sigma clips non-target pixels uniformly for each wavelength slice, replacing flagged pixels with the median of the non-target pixels in that slice.
    \item Separately sigma clips science target pixels in each S/N layer for each wavelength slice, replacing flagged pixels with the median of that S/N layer.
    \item Combines science target and non-target pixels back together into a final post-processed IFU cube.
\end{itemize}}

This algorithm works on the fully reduced level 3 data cube, which is science-ready except for the identification and removal of outlier pixels.  
By contrast, the \jdrp\ pipeline performs outlier rejection on the level 2 calibrated data products.  It uses the sampling of the same piece of the sky multiple times, through dithering, to identify outliers.  The advantage of the pipeline's method is that it works on the individual dithers, before the cube-building step in the pipeline.  It therefore should preserve spatial information; however, such an algorithm needs a very good astrometric solution, which was not initially available.  Additionally, as described above, the changing nature of the effectiveness of this step in the pipeline over time made it an unreliable method when processing IFS observations within the first year of \jwst.

We developed this algorithm in the TEMPLATES ERS collaboration to replace the outlier rejection step 
for the reasons outlined above \citep[e.g.,][]{Birkin.2023}.  We continue to use this algorithm in tandem with the current outlier rejection step in the pipeline to achieve the cleanest data products possible (see \S\ref{sec:analysis} for the motivation for this combined choice, for \jdrp\ pipeline versions 1.11.3 and later).

We now detail each part of the method.  For visualization purposes, we show as an example \jwst/NIRSpec IFS observations of SGAS1723+34, a bright, highly-magnified galaxy at $z=1.3293$ from the TEMPLATES program (Rigby et al., submitted).  Unless otherwise specified, the IFS data are reduced using calibration reference files under calibration reference data system pipeline mapping (CRDS, pmap) 1105 and \jdrp\ calibration pipeline version 1.11.3.  \tay{Finally, we set a preliminary threshold value that is 0.5--1 orders of magnitude above the highest pixel value of the science target (identified in the individual cal.fits files) in order to remove the most egregious outliers before the cube-building step in the final stage of the pipeline.}

\subsection{Generating Masks} \label{subsec:mask}

Before the outlier rejection algorithm can be run, a necessary first step is to divide all of the spaxels into those associated with the science target and those that are not (i.e., sky).  From there, we further separate the target spaxels by the spatial profile of the S/N of the target.  By generating these ``layers'' of S/N for a given target, we are able to efficiently flag and remove outliers from the data while preserving real signal from the science target.  We describe the different mask-making steps in detail in this section.

\subsubsection{Creating the Initial Science Target Pixel Mask} \label{subsubsec:initial-mask}

In order to divide the science target spaxels from the sky spaxels, we create a S/N map of the data cube based upon an identified bright spectral feature.  This step is critical, as the sky and science target spaxels are treated separately in the layered sigma clipping routine.  \tay{For this work, we use the brightest emission line (or a blend of multiple emission lines) to generate the S/N mask.  However, we note that a similar S/N map could be made by combining multiple emission lines (in the event that they trace significantly different regions of the science target) and/or creating a continuum-based S/N map by collapsing the entire cube into one map with all emission lines masked (in the event that there are no emission lines or that they trace different regions than the continuum).}

To make the S/N mask using an emission line, we take IFU slices covering a small wavelength range of $\sim$700-800 km/s centered on the brightest line of the cube (or lines, depending upon the spectral resolution of the data), which has been manually identified.  Collapsing the selected slices in the spectral direction for both the signal and uncertainty cubes \tay{from the calibration pipeline} (summing in quadrature for the uncertainty array), we generate a two-dimensional (2D) S/N map of the emission line.  From this map, we remove every pixel below a minimum threshold of S/N = 3.  This creates the first step in the target pixel mask-making step. We use this approach to making a map instead of fitting a Gaussian to the bright emission line(s) in each of the spaxels to prevent outliers in the data biasing a given emission line fit.

\begin{figure}[tb]
    \centering
    \includegraphics[width=\linewidth]{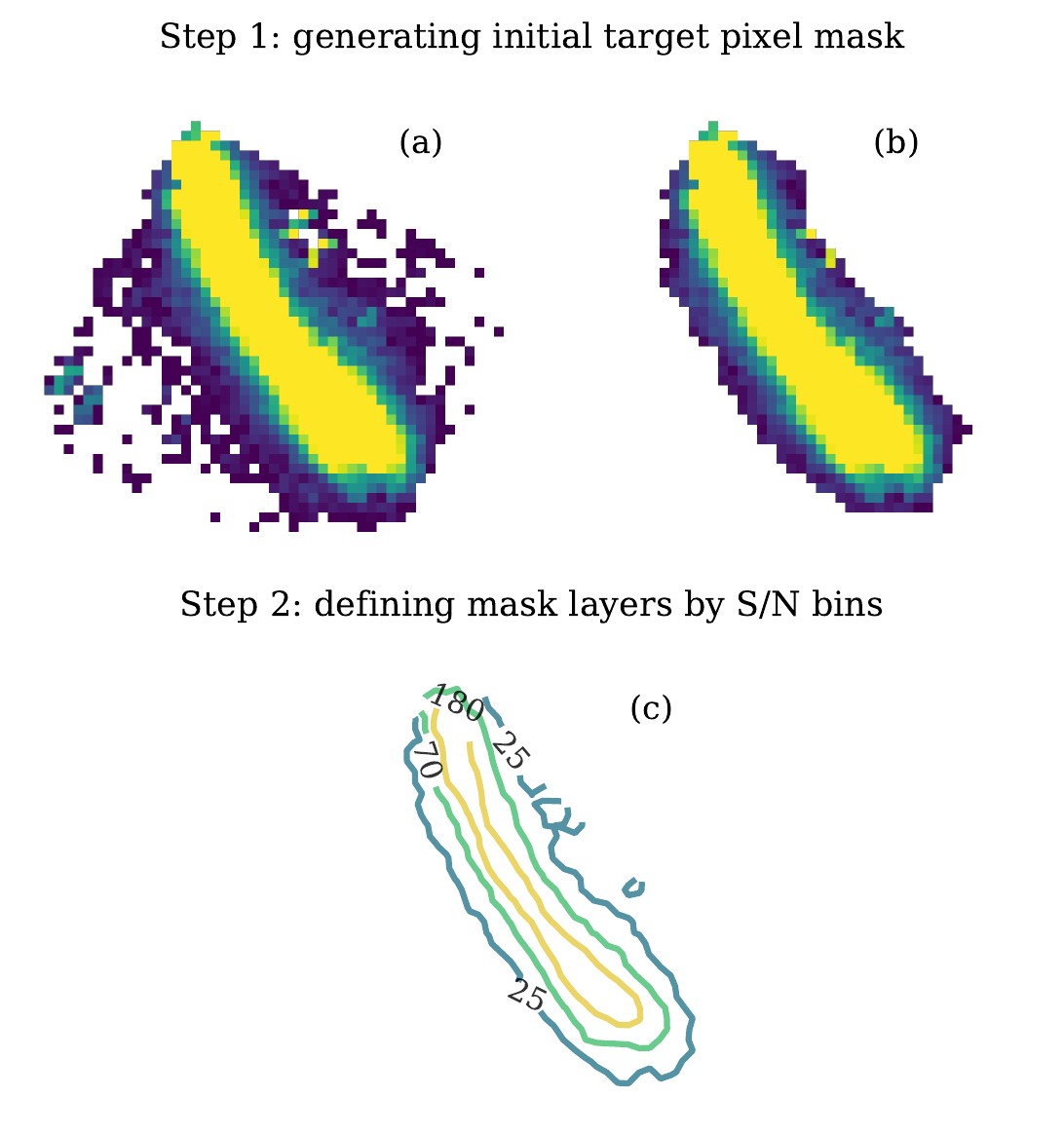}
    \caption{The two step process in making a layered target pixel mask, to use as input in the custom outlier rejection algorithm. (\textit{top}) Step 1 of the process, where (a) we generate an initial target pixel mask from the S/N map of a series of slices around a bright feature (such as an emission line), and (b) we fine-tune the mask to remove pixels that are clearly not associated with the target.  (\textit{bottom}) Step 2 of the process, where (c) we define S/N bins for our initial target pixel mask such that there are a reasonable number of pixels in each bin while not covering too large a range of S/N.  }
    \label{fig:mask}
\end{figure}

The result of the initial science target pixel mask is shown in 
Figure \ref{fig:mask}a, using the bright lensed Lyman-break galaxy SGAS1723+34 from the TEMPLATES program.  For this galaxy, we used the \oiii\ \lam5008 emission line to generate the S/N map shown.  The straightforward S/N map, created from a series of IFU slices around the bright \oiii\ feature, clearly maps out the spatial light from the target galaxy.  However, there exist additional pixels that passed the S/N threshold cut which are not obviously associated with the target.  These higher S/N sky pixels can come from surrounding sources, artifacts biasing the S/N, etc., and should be removed to create a clean science target pixel mask.  We address the fine-tuning required 
for the science target pixel mask in the following section.

\subsubsection{Fine-Tuning the Target Pixel Mask} \label{subsubsec:fine-tuning}

As evidenced by 
Figure \ref{fig:mask}a, additional steps are required to finalize the target pixel mask in order to remove sky spaxels that appear to have high S/N (either artificially high or from a nearby source). Using a pixel masking code to overlay regions on our initial target pixel mask, we visually identify spaxels that are clearly not associated with the target and remove them from the mask.  \tay{When necessary, we also visually inspected some spaxel spectra to verify whether the spaxel included light from the science target or not.} Figure \ref{fig:mask}b shows an example of the result of fine-tuning the pixel mask, where we have removed the majority of the sparse spaxels on the outskirts of the 2D map.  

As part of this process, there may be artifacts present in the summed slices that bias the S/N of certain spaxels, making them appear to have more S/N than they truly would.  This is evidenced by the few yellow and green pixels present on the upper right edge of the fine-tuned target pixel mask in Figure \ref{fig:mask}b.  The values for these pixels are not important for this step in the mask making process (as we have already identified these spaxels as associated with the target galaxy), but we will address this in the following section.  Finally, it is important to note that there may be faint signal that reaches out past the science target pixel mask made in this step.  The layered sigma clipping routine described in \sref{subsubsec:layered-clip} will not negatively affect galaxy light that faint (where the brightest line(s) are S/N $<$ 3, for example), and therefore science relevant to such diffuse light should be safe using this algorithm.\footnote{The treatment of diffuse light can be verified by checking one output of the algorithm, which tracks the pixels in each slice that are clipped.} 

After converging on a satisfactory fine-tuned science target pixel mask, we convert it into a binary mask that will be used in the algorithm (in \sref{subsec:sigma}).

\subsubsection{Splitting Mask into S/N Layers} \label{subsubsec:snr-layers}

Once satisfied with the science target pixel mask, we create the mask layers that will be used in the layered sigma clipping process.  We split the associated pixels into layers that are defined as bins of S/N.  To make these layers, we define three to four bins in the S/N map of the science target pixels.
There should be two goals when choosing the number of S/N bins for a given target: 

\begin{enumerate}
    \item Avoid too large a S/N range for a given bin to prevent real signal from being clipped in that layer \tay{(this is more critical for the lower S/N bins)}. 
    \item Have an adequate number of pixels in each bin to ensure that high-value artifacts will be clipped properly during the layered sigma clipping step \tay{(a good goal is at least 8--10 pixels in each S/N layer, so the sigma clipping performs well)}.
\end{enumerate}


Figure \ref{fig:mask}c shows an example of the S/N bins used for the same TEMPLATES lensed galaxy, with 2D contours defining the S/N bins used for the science target.  The \oiii\ emission line is incredibly bright towards the center of this lensed galaxy, with a peak S/N $>300$.  For bright extended sources whose surface brightness spans a large range in S/N across the galaxy, we recommend four S/N bins (in order to protect the brightest spaxels); for fainter sources with a smaller range, we recommend three bins.  

The number of bins used (and the S/N range spanned in each bin) will be unique to each target processed --- adjustments of the S/N ranges and bin sizes may be necessary more than once during the layered sigma clipping process (see additional discussion about this iterative adjustment in \S\ref{subsubsec:layered-clip}).  For our work, it was common to iterate \tay{a few times} between defining the target mask layers and running the layered sigma clipping routine to ensure that we achieved the best result possible.

Slight adjustments to a higher or lower S/N bin may be necessary for some individual pixels.  As referenced in the previous section, this depends upon the S/N of a given pixel and whether or not those values are driven by artifacts present in the summed IFU slices used in the original S/N map. Using the example pixels mentioned in \S\ref{subsubsec:fine-tuning}, the yellow pixels located in the upper right portion of the S/N map would fall in a higher S/N bin.  However, upon inspecting their 1D spectra, it is clear that the actual emission line in each spectrum is as faint as the spaxels around them.  Therefore, these pixels should be identified and manually moved to lower S/N bins.  

Additionally, the edges of the IFU cube have very high noise due to less coverage at the edges from dithering.  Therefore the edge pixels would be down-weighted in the 2D S/N map generated in the previous steps.  If the science target approaches or goes over the edge of the IFU field of view (as is the case for our example galaxy), these down-weighted pixels will need to be lifted to higher S/N bins to properly preserve the actual signal.
In practice, we apply a manual pixel adjustment step to several pixels for each source.  It is advantageous to go slowly through this step to ensure that the S/N bin layers, and their associated pixels, are as accurate as possible (and ideally, the pixel adjustment step is only required once per source).

Once we identify the S/N bins for a given source, we convert the pixels in each bin into Boolean masks that will be used in the layered sigma clipping routine.  Running the outlier rejection method on each of the individual layers ensures that the real signal from the science target is preserved while robustly clipping outliers.

\subsection{Custom Outlier Rejection} \label{subsec:sigma}

Here we describe the three-part custom outlier rejection algorithm, separated into three steps to allow the user to adjust the input masks as needed and check the functioning of the code along the way.   The process is divided as follows: 1) a general sigma clipping of the sky spaxels, 2) a layered sigma clipping of the target spaxels using the target pixel mask layers created in \S\ref{subsubsec:snr-layers}, and 3) combining the two separately-clipped pixel regions into one final cube.

\subsubsection{Clipping the Sky Spaxels} \label{subsubsec:clip-off-galaxy}

This step of the custom outlier rejection algorithm is straightforward.  Using the full science target pixel mask created in \S\ref{subsubsec:initial-mask}, we mask out the spaxels associated with the target.  Next, choosing a slice in the cube 
that has some amount of outliers clearly present, we identify by eye four benchmark pixels that are used to check the clipping process.  We recommend defining at least one benchmark pixel as a ``normal'' pixel, while the rest can be located on various outlier pixels in that slice.  The spectra from these pixels are plotted before and after the clipping process to verify that artifacts are being properly flagged, while ``normal'' signal is unaffected.

The clipping part of this step walks through the cube slice-by-slice, masking out the target pixels and the 
non-IFU pixels (present from to the rotation of the cube due to observing position angle).  Next, using the \texttt{sigma\_clip} function from astropy (which removes values above and below a specified standard deviation threshold), we clip the sky pixels in the slice using $\sigma$ = 5 and max\_iterations = 5.  The clipped pixels from this step are masked and replaced with the median of the unmasked sky pixels.  For users who may not want the clipped pixel values replaced, we also log the masked pixels (as a separate FITS extension) so that the user can track and/or remove the replaced values as needed.

Finally, we compare the spectra of the four benchmark pixels to check the result of the sigma clipping.  The clipped cube and associated log of clipped pixels are saved to a multi-extension FITS file, to be read in for the final step of the outlier algorithm (\S\ref{subsubsec:combining}).

\subsubsection{Layered Clipping of the Target Spaxels} \label{subsubsec:layered-clip}

This step is the most hands-on of the custom outlier rejection algorithm.
Similar to the previous step, we choose \tay{a different set of} four benchmark pixels, located across the target, for use in checking the robustness of the layered sigma clipping step in the algorithm.  When identifying the four benchmark pixels, we recommend choosing locations that cover a range in S/N for the target \tay{(from low S/N to high S/N).}\footnote{One can confirm the choice of the benchmark pixel locations by looking at slices around a particularly bright emission line or continuum; refer back to \S\ref{subsec:mask} for an example.}  In addition to the four benchmark pixels, we also compare before and after clipping of a 2D benchmark slice in the cube as additional validation that the algorithm is removing artifacts while not affecting actual signal.

At this stage, the code again runs through the cube slice-by-slice, to mask out sky pixels.  For each slice, we run through each of the S/N mask layers (defined in \S\ref{subsubsec:snr-layers}) --- setting all of the target pixels not in that layer to NaNs.  Using the \texttt{sigma\_clip} function with the same parameters as the previous step, we clip the science target pixels in the layer, masking and replacing the clipped pixels with the median of the unmasked science target pixels from the same layer. 
Clipping in this manner ensures that, for example, the higher S/N pixels are compared and clipped only with other high S/N pixels, such that real signal is preserved as much as possible while outliers are efficiently removed.  We repeat this step for each S/N mask layer.  Finally, we add the clipped layers back together as the final, clipped slice for the target pixels.  In the same manner as the previous step, we log the pixels that are clipped in each slice.  The entire process is then repeated for each slice of the cube.

\begin{figure*}[!t]
    \centering
    \includegraphics[width=\textwidth,trim = 0 30 0 10,clip]{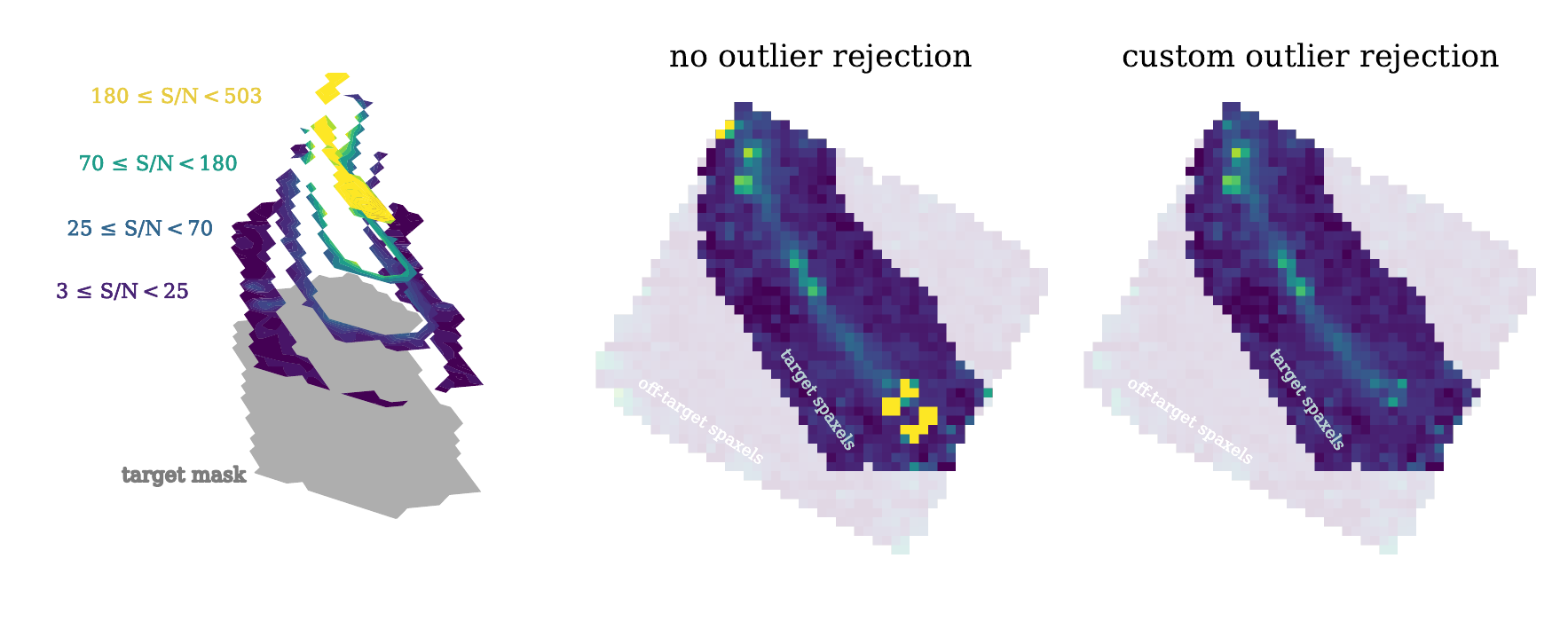}
    \caption{A benchmark slice from the reduced IFU cube for the same target galaxy, chosen to show the layered sigma clipping step of the custom outlier rejection algorithm. (\textit{left}) A view of the S/N mask layers of the target pixel mask, used to clip the target pixels in each slice as a function of mask layer in order to preserve the real signal from the science target while removing outliers. (\textit{middle}) The slice from the fully-reduced cube with no outlier detection used, showing an artifact on the bottom right corner of the science target pixels. (\textit{right}) The same slice after applying the layered sigma clipping step, showing the bright yellow pixels removed.  For both the middle and right panels, we have shaded out the sky spaxels to emphasize the science target spaxels where the layered step is applied.  }
    \label{fig:layers-slice-comparison}
\end{figure*}

\begin{figure*}[t]
    \centering
    \includegraphics[width=\textwidth]{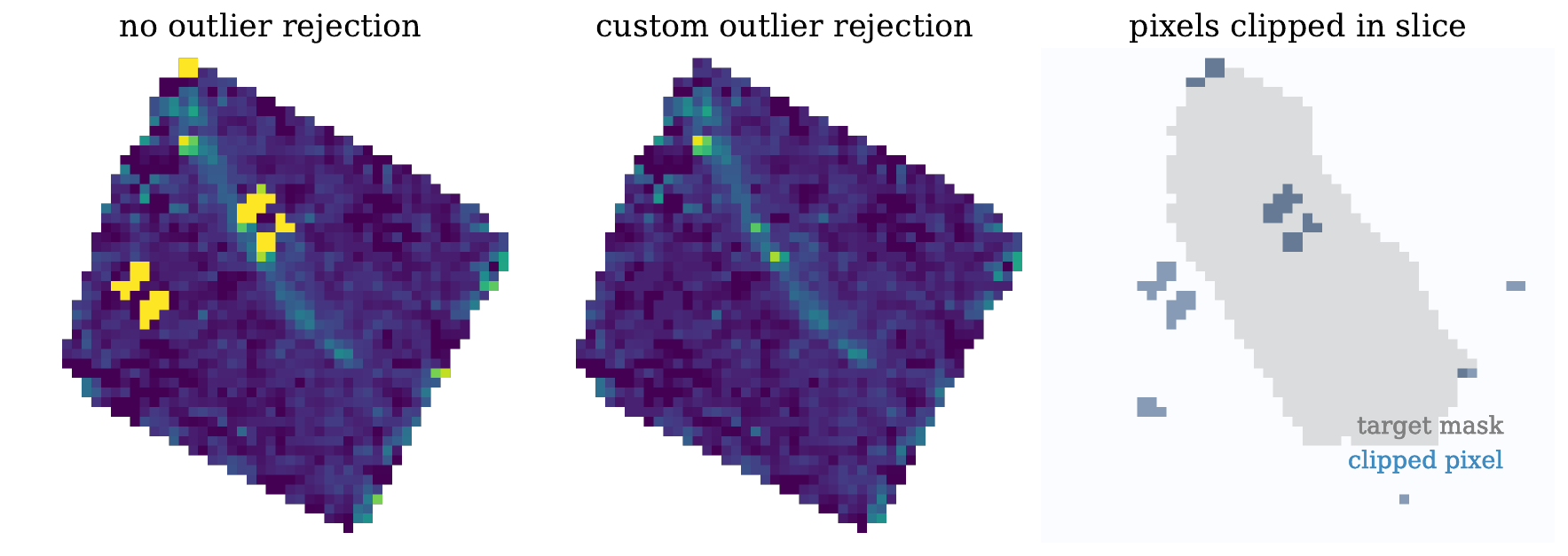}
    \caption{A view of a different benchmark slice in the reduced IFU cube for the same target galaxy, chosen to highlight the efficacy of the custom outlier rejection algorithm.  (\textit{left}) The slice from the fully-reduced cube with no outlier detection used, showing artifacts in the form of bright yellow pixels.  (\textit{middle}) The same slice after applying the custom outlier rejection algorithm, showing the artifacts removed. (\textit{right}) The pixel logging extension included in the code, to track which pixels in each slice have been clipped by this method (blue) with the target pixel mask underlaid (grey). }
    \label{fig:slice-comparison}
\end{figure*}

Figure \ref{fig:layers-slice-comparison} shows an example of the result of the layered clipping step, with the pre-defined S/N pixel mask layers from \S\ref{subsubsec:snr-layers} shown in the left panel.  This visualization of the S/N layers highlights the layered clipping effect, where in each slice the code runs through the sigma clipping of each layer individually.  The middle and right panels of Figure \ref{fig:layers-slice-comparison} show the result for a benchmark IFU slice, chosen to illustrate the layered sigma clipping step on an area of the target that covers a range of S/N (or, more than one S/N mask layer).

\begin{figure*}[!th]
    \centering
    \includegraphics[width=0.98\linewidth]{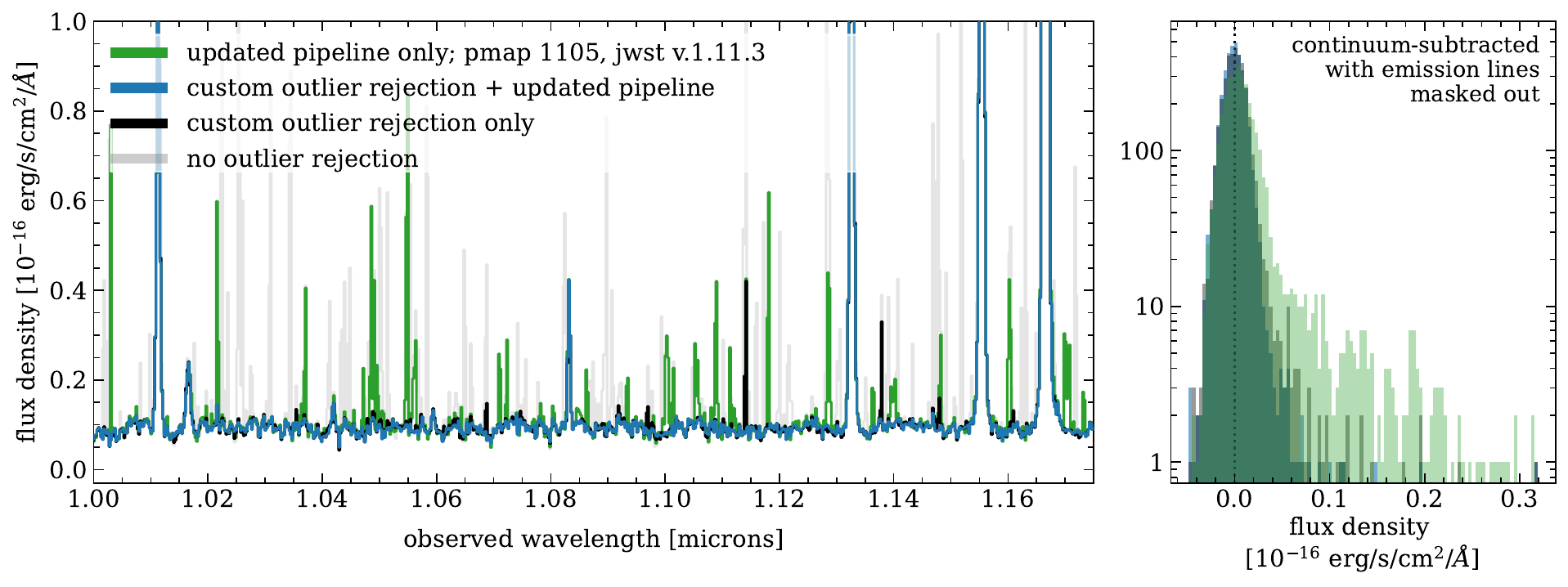}
    \caption{
    (\textit{left}) A comparison of our custom outlier rejection algorithm (black), the updated outlier detection step in the \tay{version 1.11.3} \jdrp\ pipeline (green), and a combination of both methods (blue) for the G140H grating for our example galaxy.
    The spectra shown are spatially integrated from the target spaxels for the different outlier rejection codes.  We have arbitrarily chosen this representative spectral region to highlight the leftover artifacts present in the data after each outlier rejection method.   Additionally, as a reference for the utility of both our algorithm and the updated \jdrp\ pipeline in removing most artifacts, underlaid in this panel we have included an example using no outlier rejection (faint grey).
    (\textit{right})
    A histogram of the flux density of the three co-added spectra (green, blue, black) demonstrating the overall number of artifacts leftover from the different algorithms.  The spectra used in this histogram have been continuum-subtracted with the emission lines masked out in order to highlight the artifact spikes still present in the data. 
    }
    \label{fig:comparing-clipping-outlier}
\end{figure*}

The result of the second step in the algorithm shows a clean removal of outliers while preserving the signal from the science target.  
\tay{While this step does spatially interpolate to replace the flagged value, we argue that this method of replacing the flagged science target pixels using the S/N layers, for each wavelength slice, is a reasonable approximation due to the definition of the S/N layers themselves.  They are defined such that the pixels flagged in each layer, for a given wavelength slice, are compared with (and replaced by) only those with similar S/N values (and therefore relatively similar flux densities).  Additionally, in practice we have seen outliers impact at most only two (and at rare instances, three) consecutive wavelength slices for a given spaxel, while spectral features of the science target such as emission lines generally span many more wavelength slices.  Therefore, if this algorithm's approach happens to poorly replace flagged signal for a given spaxel, it should be evident with little effort.  
Additionally, for spatially-integrated science using IFS data, if there are missing values then the integrated flux measurements will be artificially low (requiring some method of correcting for the percentage of flagged pixels).
However, if the user prefers to flag but not replace pixels, we include the log of clipped pixels (for each wavelength slice) as part of the output of the algorithm for this purpose.}

Upon completion of this step, we \tay{visually} compare the four benchmark pixels to measure how well the layered sigma clipping procedure performed.  
These comparisons and checks are vital for this step of the custom outlier rejection algorithm.  Some adjustments may be necessary between the layered sigma clipping step and in the previous step where we define the S/N mask layers used (\sref{subsubsec:snr-layers}).  If the code appears 
to be overzealous in clipping real light, or if it is missing obviously false signal, adjusting the S/N bin ranges described in \S\ref{subsubsec:snr-layers} will be necessary.  

The clipped target cube and associated log of clipped pixels are saved to a multi-extension FITS file.

\subsubsection{Creating the Final Cubes} \label{subsubsec:combining}
For the final step of the custom outlier rejection method, we combine the output from the previous two steps.  In short, we combine the slices from each of the previous steps to create a final complete slice, repeating this process for the full cube.  We keep this step separate from the previous steps to enable validation checks at each step in the algorithm.  As part of this, we compare the original cube with our post-processed cube at various slices where we had previously identified the presence of outliers.  This inspection is important to ensure that the final product is appropriately science ready, and that real signal has been properly preserved throughout the algorithm.  

Figure \ref{fig:slice-comparison} shows such a comparison for another benchmark IFU slice from our example galaxy, chosen to highlight artifacts both on and off of the target galaxy.  The right panel shows the same slice, but from the aforementioned log of clipped pixels, showing the pixels in the benchmark slice that the code flagged in both of the previous steps.  We include the target pixel mask in this panel to denote the location of flagged pixels on and off of the science target.

The final clipped cube and associated  log of clipped pixels are saved to a multi-extension FITS file, and are the science-ready data products produced by the algorithm.

\section{Comparisons with\\ the Existing Pipeline} \label{sec:analysis}

In this work, we have detailed a custom outlier rejection algorithm that we originally designed to replace the \jdrp\ outlier detection step.  As previously described, this was motivated in large part due to the pipeline (versions 1.10.2 and earlier) aggressively removing outliers such that it was incorrectly flagging and removing signal from bright emission lines in our science targets, resulting in stunted and oddly-shaped emission line profiles.  As of \jdrp\ pipeline version 1.11.3, the outlier detection step has been improved such that it no longer flags and removes real signal from our science targets.  However, artifacts still remain in the final cubes --- they require additional processing to remove.  

We show the utility of the outlier rejection methods and the leftover outliers present in each data cube in Figure \ref{fig:comparing-clipping-outlier}.  In the left panel of Figure \ref{fig:comparing-clipping-outlier}, we show spatially integrated 1D spectra for both the \tay{version 1.11.3} \jdrp\ pipeline (green) and our outlier rejection algorithm (black) as well as a combination of the two (blue).  For comparison, an example of IFS with no outlier rejection applied is underplotted (faint grey) to highlight how both methods successfully flag and remove \textit{most} of the outliers in the IFS data.  The effect of our algorithm is striking.  For the representative spectral region, we find that the \tay{version 1.11.3} pipeline's outlier detection step produces $\sim$ 25 leftover artifacts.\footnote{To quantify the leftover artifacts, we count the peaks of the noise spikes but note that some artifacts span more than one wavelength slice.}  
In contrast, our algorithm produces $\sim$5 leftover artifacts in this same spectral region. Thus, our algorithm results in one fifth as many residual artifacts as the updated \tay{version 1.11.3} \jdrp\ pipeline.\footnote{\tay{We utilize the default parameters in the pipeline's outlier detection step.  Detailed testing of tweaking the input parameters for that step and rerunning the pipeline iteratively may help reduce the number of artifacts that remain in the data.}}  

Using the combination of both methods --- where first we use the \tay{version 1.11.3} pipeline's outlier detection step in stage 3 and then apply our algorithm on the final 3D cube --- yields only 1 artifact remaining in the spectral region shown.  Thus, combining both approaches produces data cubes with the cleanest removal of artifacts.  

We visualize the total number of leftover artifacts in the right panel of Figure \ref{fig:comparing-clipping-outlier}, where we show histograms of the flux density of the spatially integrated 1D spectra that are shown in the left panel.  To make the comparison of leftover artifacts from each method easier to quantify, we have subtracted the continuum and masked out all emission lines in each spectrum.  For clarity, we have not included the IFS data with no outlier rejection in this panel.  The \tay{version 1.11.3} pipeline's outlier detection step (green) very clearly has the most remaining artifacts of the two methods, with many flux density values biasing towards larger positive values.  By comparison, our algorithm (black) has remaining artifacts that exist at similar high values, but there are significantly fewer present.  The differences between the two methods verify the results found in the left panel, where our algorithm catches and removes a much larger number of leftover artifacts than the \tay{version 1.11.3} pipeline itself.  However, the combination of the two methods (blue) shows the tightest result, with very few values above 10$^{-17}$ erg s$^{-1}$ cm$^{-2}$ $\AA^{-1}$ for the science target.

From these results, we conclude that the combination of these two methods performs better than either method alone.  This combined approach has the added benefit that the \jdrp\ outlier detection step works on on the individual dithers, while our algorithm works on the final, drizzled three-dimensional cube.  
In principle, the pipeline's approach is preferable, since it removes outliers from the individual dithers before combining; with a sufficient number of dithers, the flux density of a given sky spaxel can be determined from the dithers not affected by an artifact.  The uncertainty in that flux density will be greater since the removal of the outlier effectively lowers the total integration time for that pixel, but there is no interpolation of any kind.  By contrast, as our algorithm works on the final drizzled spectral cube, it necessarily interpolates spatially, replacing the flagged outliers with the median of spaxels at similar S/N and flux density levels.  This philosophical difference in approach is why, for the TEMPLATES ERS program, we chose to use the pipeline's outlier detection first (which does not interpolate) and then use our algorithm on the outliers that the pipeline still misses (see Rigby et al., submitted, for more details).

\section{Summary \& Release of Code} \label{sec:conclude}

We present a custom outlier rejection method for use with \jwst/NIRSpec IFS data.  This custom method employs a layered sigma clipping approach that uses the spatial profile S/N of the science target, which preserves the signal of the science target while efficiently and robustly removing outliers, primarily due to cosmic rays.  
Along with this paper, we release the associated algorithm code as Jupyter notebooks for use by the community.  The repository for the code can be found at \href{https://github.com/aibhleog/baryon-sweep}{github.com/aibhleog/baryon-sweep} (DOI: 10.5281/zenodo.8377532).  \tay{This algorithm requires some manual customization --- however, the results produce one of the lowest number of remaining outliers of which we are aware, therefore we argue that the final product makes the effort worthwhile.}

We developed this algorithm originally as a replacement of a critical step in the \jdrp\ pipeline. 
We compare to current \tay{version 1.11.3} \jdrp\ pipeline step and find that our algorithm results in one fifth as many residual artifacts as the \jdrp\ pipeline.  Further, we find that running the updated outlier detection in \tay{version 1.11.3} of the \jdrp\ pipeline, and then running our algorithm, produces data that are nearly completely cleaned of outliers.  We use this combination for our IFS data in the TEMPLATES DD-ERS program.  

Finally, an alternative approach to this algorithm could include applying the layered sigma clipping method to the individual 2D dithers (therefore working with the data before the cube-building stage, like the \jdrp\ pipeline's step).  Additionally, the mask-making step of this algorithm could be altered to instead use Voronoi binning, Fourier transforms, and other such methods to separate target spaxels from non-target (or sky) spaxels.  Depending upon the method used (and whether a S/N map or a flux map is utilized), such alternative methods could be more automated than the mask-making method used in this algorithm.  \tay{Future improvements to the algorithm could include two-dimensional spline interpolation at each S/N layer instead of replacing the clipped values with the median of the S/N layer.}


\begin{acknowledgments}

This work is based on observations made with the NASA/ESA/CSA \jwst. The data were obtained from the Mikulski Archive for Space Telescopes at the Space Telescope Science Institute, which is operated by the Association of Universities for Research in Astronomy, Inc., under NASA contract NAS 5-03127 for \jwst. 
We are grateful for the collective contributions of the roughly 20,000 individuals around the world who designed, built, tested, commissioned, and operate \jwst.
%
These observations are associated with \jwst\ program \# 1355. 
Support for \jwst\ program \# 1355 was provided by NASA through a grant from the Space Telescope Science Institute, which is operated by the Association of Universities for Research in Astronomy, Inc., under NASA contract NAS 5-03127.
TAH is supported by an appointment to the NASA Postdoctoral Program (NPP) at NASA Goddard Space Flight Center, administered by Oak Ridge Associated Universities under contract with NASA. 
BW acknowledges support from NASA under award number 80GSFC21M0002.
JEB \& GMO acknowledges generous support from the Texas A\&M University and the George P. and Cynthia Woods Institute for Fundamental Physics and Astronomy.

\end{acknowledgments}

%

\vspace{5mm}
\facilities{\textit{JWST}(NIRSpec)}


\software{\jdrp\ Calibration Pipeline \citep{Bushouse.2023},
          astropy \citep{astropy:2013,astropy:2018,astropy:2022},  
          scipy \citep{2020SciPy-NMeth},
          matplotlib \citep{matplotlib},
          stenv (\href{https://stenv.readthedocs.io/en/latest/}{https://stenv.readthedocs.io/en/latest/}),
          pandas \citep{reback2020pandas}
          }



\appendix

\begin{figure}[!th]
    \centering
    \includegraphics[width=\linewidth]{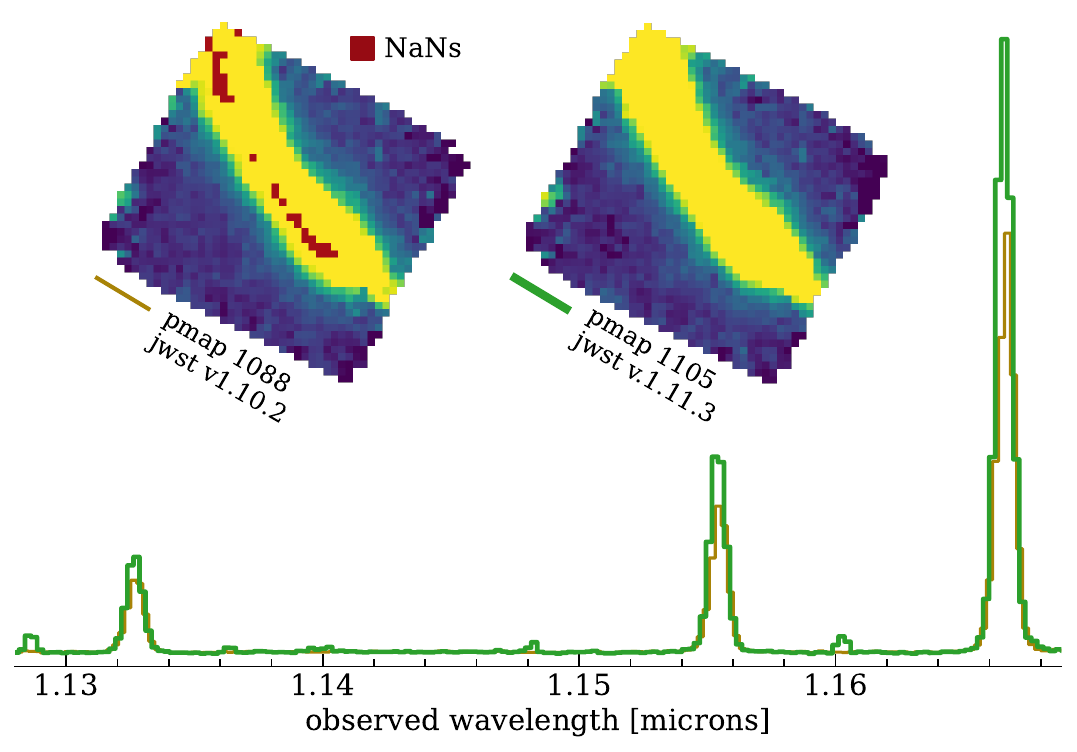}
    \caption{A visualization of an issue in a critical step in the \jdrp\ pipeline that motivated a large part of this work.  Plotted is a zoom-in on emission lines in a spatially-integrated spectrum of science target SGAS1723+34, showing the difference between the outlier detection step in the \jdrp\ pipeline software before (gold, \tay{version 1.10.2}) and after (green, \tay{version 1.11.3}) the recent (c.\ August 2023) update.  Insets above the spectra show the same effect but in a 2D wavelength slice of the cube, centered on an emission line where galaxy signal was initially improperly clipped (red pixels).   Both the 2D and 1D data show that the updated \tay{version 1.11.3} \jdrp\ outlier detection step no longer incorrectly flags and removes actual signal in IFS data, a welcome improvement. }
    \label{fig:old-new-outlier}
\end{figure}

\section{\tay{comparison of the original \& updated \jdrp\ outlier detection step}}

During the first year of observations, the \jdrp\ pipeline's outlier detection step did not work, due to an initial coarse instrument model that later was found to be removing not only outliers present in the NIRSpec IFS data, but also removing real flux from the science target \citep[\jdrp\ calibration pipeline versions 1.10.2 and earlier; e.g.,][among others]{Perna.2023,Veilleux.2023,Marshall.2023,Vayner.2023}. 

Figure \ref{fig:old-new-outlier} illustrates this effect for our example science target in both a 2D wavelength slice (centered on the \oiii\ \lam5008 emission line) and in spatially integrated 1D spectrum zoomed in to show the \hb\ \lam4864 and \oiii\ \lam4960,5008 emission lines.  \tay{The data shown are from \jdrp\ pipeline versions 1.10.2 and 1.11.3, respectively.}  In both views, the effect of this overly aggressive outlier detection in the \tay{version 1.10.2} \jdrp\ is clearly visible --- from the red pixels denoting NaNs in the left 2D wavelength slice to the stunted and oddly-shaped emission line profiles in the gold-colored spectrum.

As previously noted, the \jdrp\ pipeline's outlier detection step has recently (c.\ August 2023) been improved such that it no longer removes real signal from science targets \tay{(versions 1.11.3 and later)}.  Results from the updated \tay{version 1.11.3} \jdrp\ pipeline's outlier detection step are shown in Figure \ref{fig:old-new-outlier}, in both a 2D wavelength slice 
(right) and the green-colored spatially integrated 1D spectrum, where the emission line profiles have the correct line strengths and line profiles.  

In additional testing of the updated pipeline outlier detection step for all of the TEMPLATES NIRSpec IFS data, we have found consistent results indicating that real signal is preserved with the newly updated pipeline --- a significant improvement.  \tay{However, the updated version 1.11.3 \jdrp\ pipeline outlier detection step does not remove all outliers (as evidenced in \sref{sec:analysis}).}


\bibliography{ref}{}
\bibliographystyle{aasjournal}



\end{document}